\documentclass[tightenlines,aps,twocolumn,showpacs,floatfix]{revtex4}
\usepackage{amsmath,latexsym,amsfonts,graphicx,bm}
\newcommand{\p}[1]{b_{#1}}
\newcommand{\q}[1]{s_{#1}}

\begin{document}      
%

%


%
\title{
\[ \vspace{-2cm} \]
\noindent\hfill\hbox{\rm  SLAC-PUB-9296} \vskip 1pt
\noindent\hfill\hbox{\rm hep-ph/0207201} \vskip 10pt
The Asymptotic Expansion of Lattice Loop Integrals Around
the Continuum Limit
}      
\author{Thomas Becher and Kirill Melnikov}      
    
\affiliation{Stanford Linear Accelerator Center, Stanford University,
Stanford, CA 94309, U.S.A.}         

\begin{abstract}    
We present a method of computing any one-loop integral in lattice
perturbation theory by systematically expanding around its
continuum limit.  At {\it any} order in the expansion in the lattice
spacing, the result can be written as a sum of continuum loop
integrals in analytic regularization and a few genuine lattice
integrals (``master integrals''). These lattice master integrals are
independent of external momenta and masses and can be computed
numerically.  At the one-loop level, there are four master integrals
in a theory with only bosonic fields, seven in HQET and sixteen in QED
or QCD with Wilson fermions.
\end{abstract}      
\maketitle     

\section{Introduction}

 Lattice regularization provides a non-perturbative formulation of
quantum field theories and allows for the approximate numerical evaluation
of the associated path integral \cite{Rothe:kp}. 
Nevertheless, perturbative calculations with lattice regularization
are often necessary, particularly when the problem at hand contains
several disparate scales. In such situations it is sometimes possible
to separate perturbative short distance physics from non-perturbative
long distance physics.  This separation depends on the regularization
prescription, and in order to combine the perturbative results for the
short distance part with the non-perturbative evaluation of the long
distance matrix elements, both must be computed within the same
regularization scheme. A typical example is the weak decay of a
hadron. The physics associated with the high scale $M_W$ can be
treated perturbatively by absorbing it into the Wilson coefficients of
a low energy effective Lagrangian. The low energy physics is then
obtained by evaluating the matrix elements of the operators that
induce the decay. To match the lattice results for the matrix elements
with the Wilson coefficients, typically obtained in dimensional
regularization, a perturbative calculation in lattice regularization
must be performed. Other important applications of lattice
perturbation theory include the construction of actions with smaller
discretization errors \cite{Symanzik:1983dc}, and the extraction of
quark masses \cite{Sint:1998iq} and the strong coupling constant
\cite{Luscher:1995nr,Luscher:1995np,Alles:1996cy} from lattice
simulations.

It is well known that perturbative calculations with lattice
regularization are difficult. The propagators and interaction vertices
in lattice regularization are much more complicated than their
continuum counterparts; lattice regularization also badly violates
Lorentz invariance. Consequently, the standard set of tools for
one-loop calculations in continuum perturbation theory does not appear
to be useful in lattice calculations.  In the standard approach to
lattice perturbation theory the relevant loop integrals are therefore
evaluated numerically, which has several drawbacks: a) the amount of
numerical computations necessary for realistic calculations is huge;
b) cancellations between individual diagrams can render numerical
results unstable; c) the continuum limit, i.e. the limit in which the
inverse lattice spacing becomes much larger than external momenta and
masses, has to be taken numerically as well. There exist a number of
techniques to both reduce the amount and enhance the precision of the
numerical integrations involved. These methods rely on Reisz's power
counting theorem \cite{Reisz:1987da} and use momentum subtractions to
split the loop integrals into lattice tadpole contributions and a
remainder whose continuum limit can be taken naively. The number of
lattice tadpole-integrals is then reduced by exploiting relations
among them \cite{Caracciolo:1991cp,Burgio:1996ji,Luscher:1995zz} and
techniques for the precise numerical evaluation of such integrals have
been developed \cite{Luscher:1995zz}.

In this paper we demonstrate that a completely different approach to
lattice perturbation theory is possible if advanced methods of
continuum perturbation theory are applied to lattice calculations.  We
first show how to construct the systematic expansion of any lattice
integral around its continuum limit by employing the technique of
asymptotic expansions \cite{Smirnov:pj} developed for continuum loop
integrals. After the expansion, the original lattice integral is
expressed as a sum of two different contributions: continuum one-loop
integrals that are regularized by means of analytic regularization and
massless lattice tadpole-integrals.  We then use integration-by-parts
identities \cite{ibp} to systematically reduce the number of 
tadpole-integrals to an absolute minimum.  Those remaining are called master
integrals and are evaluated numerically; their number depends on
the theory and varies from four in any theory with only bosonic fields
to sixteen in QED or QCD with Wilson fermions.

The advantage of this approach is that the one-loop lattice 
tadpole-integrals relevant for any conceivable calculation can be expressed in
terms of master integrals in a process independent way. Therefore,
once these relations are established and the master integrals are
computed numerically, {\it perturbation theory on the lattice reduces
to perturbation theory in the continuum}.

Our paper is organized as follows. After presenting our notation, we
illustrate the method by applying it to massive tadpole-integrals.
Although this example is very simple, it exhibits all the basic
features of the method. We then study Feynman diagrams that involve
static and Wilson fermions. In both cases the loop integrals can be
expanded around their continuum limit in exactly the same way as the
bosonic integrals.  However, the comparatively more complicated form
of the propagator for Wilson fermions leads to a larger number of
master lattice integrals. Finally, we use the technique to rederive
the known results for the gluon self-energy and the static and Wilson
fermion two-point functions at one loop order in perturbation theory.

\section{Preliminaries\label{sec:notation}}

In addition to particle masses and momenta, the result of a
calculation in lattice perturbation theory depends on the lattice
spacing $a$.  It is convenient to define dimensionless quantities
by multiplying all momenta and masses by the lattice spacing so that
$m=a\,m_{\rm phys}$, $p=a\, p_{\rm phys}$, etc. We use
the standard lattice notation
\begin{equation}
\widehat{p}^2=\sum_{\mu=1}^d \widehat{p}_\mu^2,~~~~~~
\widehat{p}_\mu = 2 \sin\frac{p_\mu}{2}\, ,
\end{equation}
throughout the paper. The massive bosonic propagator on the lattice is
\begin{equation}
G_B(k) = \frac{1}{(\widehat{k}^2 + m^2)}.
\label{prboson}
\end{equation}
Loop integrals contain products of these propagators integrated
over the Brillouin zone; a typical loop integral has the form
\begin{equation}
\int \limits_{-\pi}^{\pi}
\frac{d^d k}{(2\pi)^d} \frac{1}{(\widehat{k}^2
+ m^2)}\,\frac{1}{((\widehat{p+k})^2 + m^2)}\,,
\label{e1}
\end{equation}
where $p$ is the external momentum and $d=4$ is the space-time
dimension.

Our goal is to construct a procedure for expanding lattice integrals
around their continuum limit. In terms of dimensionless quantities,
this limit corresponds to all external momenta and masses of the
particles becoming small. We begin by mapping the integration region
in Eq.(\ref{e1}) to an infinite volume and define new integration
variables $\eta_\mu$,
\begin{equation}
\eta_\mu = \tan(k_\mu/2).
\label{beq}
\end{equation}
In terms of the new variables, the loop integrations in Eq.(\ref{e1})
range from $-\infty$ to $+\infty$. It turns out to be convenient to
introduce variables similar to Eq.(\ref{beq}) also for external
momenta and to rescale $m$ to $\tilde m=m/2$. The self-energy integral in
Eq.(\ref{e1}) then becomes
\begin{eqnarray}
&& \frac{1}{4^2\,\pi^d} \int \limits_{-\infty}^{\infty}
\prod_{i=1}^d
\frac{d\eta_i}{(1+\eta_i^2)}
\left [\tilde m^2+\sum \limits_{i=1}^d \frac{\eta_i^2}{(1+\eta_i^2)} \right ]^{-1}
\nonumber \\
&& \times
\left [ \tilde m^2+\sum \limits_{i=1}^d 
\frac{(\rho_i+\eta_i)^2}{(1+\eta_i^2)(1+\rho_i^2)}
\right ]^{-1}, 
\label{eq1a}
\end{eqnarray}
where $\rho_i = \tan(p_i/2)$.

The representation of lattice loop integrals as in
Eq.(\ref{eq1a}) is similar in form to continuum loop
integrals, albeit with unconventional propagators.  The form of these
propagators makes it impossible to apply the basic continuum
techniques for performing loop calculations, such as Feynman
parameterization and Passarino-Veltman reduction. However, we will
show in the next section that more advanced techniques, such as recurrence
relations and asymptotic expansions, {\em are} applicable.

\section{Expansion around the continuum limit}

We demonstrate our method by evaluating the massive bosonic
tadpole-integral. Although this is the simplest  loop integral,
the treatment of integrals with external momenta and different masses
does not pose any additional difficulty; the part of these
integrals that has nontrivial dependence on the masses and momenta, 
can be obtained by evaluating continuum one loop integrals.  Using the 
variables defined above, the massive lattice tadpole-integral can be written 
as
\begin{equation}
G(\tilde m) = \frac{1}{4\,\pi^d}
\int \limits_{-\infty}^{\infty} \prod_{i=1}^d
\frac{d\eta_i}{(1+\eta_i^2)} \left [  \tilde m^2+
D_B(\eta) \right ]^{-1},
\label{eq2}
\end{equation}
where 
\begin{equation}
D_B(\eta) = \mbox{$\sum \limits_{i=1}^d$}
\frac{(\eta_i)^2}{(1+\eta_i^2)}.
\end{equation}

We are interested in the continuum limit of Eq.(\ref{eq2}), $\tilde m
\to 0$.  $G(\tilde m)$ therefore depends on a small
parameter and this fact can be used to simplify the calculation of the
integral in Eq.(\ref{eq2}). Unfortunately, a straightforward Taylor
expansion of the integrand in $\tilde m$ is not possible, since
$G(\tilde m)$ is not an analytic function of $\tilde m$.

Traditionally, lattice loop integrals with external momenta are
simplified using Reisz's power counting theorem
\cite{Reisz:1987da}. In this approach, a number of momentum
subtractions are performed on the integral. These subtractions split
the integral into two parts: a polynomial in the external momenta with
tadpole-integrals as coefficients, and a remainder that depends
non-trivially on the external momenta, but whose continuum limit can
be taken naively. When applied to theories with massless fields, this
method produces infrared divergent tadpole-integrals, and makes it
necessary to introduce an infrared regulator at intermediate stages of
the calculation. These infrared divergences illustrate that the loop
integrals are not analytic functions of the small momenta and masses;
this is reflected in the appearance of terms such as $\log( \tilde m)$.

As is well known from continuum perturbation theory, the fact that the
Taylor expansion of the Feynman integral in a small parameter does not
commute with the loop integration is not an obstacle for constructing
a highly practical procedure to perform the expansion {\it before the
integration}; this technique is known as the asymptotic expansion of
loop integrals \cite{Smirnov:pj}. To apply this procedure to lattice
loop integrals we must introduce an additional regulator into the
integral in Eq.(\ref{eq2}).  We use analytic regularization, and replace
the integral $G(\tilde m)$ by
\begin{equation}
G(\tilde m) = \frac{1}{4\,\pi^d}
\int \limits_{-\infty}^{\infty} \prod_{i=1}^d
\frac{d\eta_i}{(1+\eta_i^2)} \left [ \tilde m^2+
D_B(\eta) \right ]^{-1-\delta}.
\label{eq3}
\end{equation}
After the regulator is introduced, the expansion of $G(\tilde m)$
in $\tilde m$ can be constructed; the limit
$\delta\rightarrow 0$ can  be taken at the end.  The result for
$G(\tilde m)$ is obtained as a sum of two contributions:
\begin{equation}
G(\tilde m) = G_{\rm soft} (\tilde m) + G_{\rm hard}(\tilde m).
\label{eq4}
\end{equation}
The soft and hard contributions are calculated by applying
the following procedure to the integrand in Eq.(\ref{eq3}):
\begin{itemize}
\item Soft: assume that all the components of the
loop momentum $\eta$  are small, $\eta_i \sim \tilde m \ll 1 $.
Perform the Taylor expansion of  the integrand in Eq.(\ref{eq3}) in
the small quantities $\eta_i$ and $\tilde m$. 
The expansion coefficients in this region
are standard continuum one loop integrals, regularized analytically.
Note that {\it no} restriction on the integration region is
introduced.
\item Hard: assume that all the components of the loop momentum are large,
$\eta_i \sim 1 \gg \tilde m$ and Taylor expand the integrand in $\tilde m$. 
The expansion coefficients are given by the functions \makebox{$H(\{a_i\};n)$},
defined in Eq.(\ref{basishard}) below. These functions are
related by algebraic and integration-by-parts identities.
\end{itemize}

We note that the two regions discussed above do not account for all
the possible scalings of the components of the loop momentum. A
potential contribution arises from regions where some components of the
loop momentum are soft and the other are hard. Following the above
logic, the integrand of Eq.(\ref{eq3}) should be expanded in the small
quantities; it is easy to see that in this case the analytic
regularization as introduced in Eq.(\ref{eq3}) {\it does not} fully
regulate the resulting expressions.  The contributions of these mixed
regions are set to zero. To justify this prescription we consider
another possible choice of the regulator. We regulate the
loop integrals by including a factor $(\sin^2(k_i/2))^\epsilon$ in the
measure for each component of the loop momenta. The disadvantage of
this regulator is that the resulting soft parts are more difficult to
calculate than in analytic regularization. Its advantage is that it
fully regulates the mixed regions, where some loop momentum components
are small and others are large.  The integrals occurring in the mixed
region are scaleless and therefore vanish.  We now introduce both the
sine-function and the analytic regulator simultaneously to show that
the mixed regions do not contribute. Since the sine-regulator does not
contain any unregulated regions, the original integral is recovered by
the prescription
\begin{equation}
G(M)=
\lim_{\delta\rightarrow 0}\,\lim_{\epsilon\rightarrow 0}
G(M,\epsilon,\delta).
\end{equation}
In this prescription the mixed regions do not contribute. After
showing that for non-zero $\delta$ both the hard and the soft parts
are analytic functions of $\epsilon$ at $\epsilon=0$, we can take the
limit $\epsilon\rightarrow 0$ at the level of the integrand and thus
recover the integrals in analytic regularization. To prove
analyticity of the hard part, we use the relations between the
integrals that appear in the hard part to express  them through
 convergent master integrals
and verify that there are no $1/\epsilon$ singularities in the
relations of any hard integral to the master integrals. We can
therefore drop the sine-function regulator and ignore the
contributions of the mixed regions. The analytically regularized
integral is then recovered and its expansion is constructed according
to the rules given after Eq.(\ref{eq4}). We now discuss how the
soft and the hard contributions to the massive tadpole-integral are
calculated.

\subsection{Soft part of the integrals}

The soft part of $G(\tilde m)$ is easy to obtain.  After expanding the
integrand, we arrive at analytically regularized {\em continuum} one
loop integrals. Higher orders in the expansion introduce a large
number of strongly divergent integrals with non-covariant numerators;
a consequence of broken Lorentz invariance on the lattice.  This is
not an obstacle, since all such integrals can be easily evaluated
using the identity
\begin{eqnarray}
&& \int {\rm d}^d \eta\, 
\frac{1}{\left(\eta^2+\tilde m^2\right)^\alpha}\,\prod \limits_{i=1}^d (\eta_i^2)^{a_i}
={(\tilde m)}^{d -2\alpha+ 2\sum_i a_i}\, 
\nonumber \\
&& \times \frac{\Gamma(\alpha-d/2 - \sum_i
a_i)}{\Gamma(\alpha)}\,
  \prod_i  \Gamma(\frac{1}{2} + a_i) \, .
\label{eq:loopInt}
\end{eqnarray}
The calculation of the soft part of integrals that depend on external
momenta or several distinct masses is equally simple: this part is
always given by continuum integrals in analytic regularization and can
be brought to the form of Eq.(\ref{eq:loopInt}) using standard
methods, such as Feynman parameters.

The result for the soft part of the massive
tadpole-integral is
\begin{multline}
4\pi^2 G_{\rm soft}(\tilde m)
= -\tilde m^2\,\left( \frac{1}{\delta } +
     1 - 2\,\log (\tilde m) \right) \\
+ \tilde m^4\,\left(
     \frac{1}{2\,\delta } + \frac{3}{4} - \log
     (\tilde m) \right)+O(\tilde m^6).
\end{multline}

\subsection{Hard part of the integrals}

We now consider the hard contribution to $G(\tilde m)$.  Recall that
the hard part is obtained by expanding Eq.(\ref{eq3}) in a Taylor series
in $\tilde m$; the resulting expression is
\begin{multline*}
4\pi^4\,G(\tilde m)_{\rm hard}=H(\{ 1,1,1,1\},1 ) \\- 
\tilde m^2\,\left( 1 + \delta \right) \, H(\{ 1,1,1,1\},2 )\\ +
\frac{\tilde m^4}{2}\,\left( 1 + \delta\right)\left(2+\delta \right)\,
H(\{1,1,1,1\},3)+\dots \, .
\end{multline*}
The functions $H(\{a_i\};n)$ are defined by
\begin{equation}\label{basishard}
H(\{a_i\};n)=  \int \limits_{-\infty}^{\infty}
\prod_{i=1}^d \frac{d\eta_i}{(1+\eta_i^2)^{a_i}}
\left [D_B(\eta) \right ]^{-n-\delta} \, ,
\end{equation}
where $a_i$ and $n$ are integers. In a theory with only bosonic
propagators, the functions $H(\{a_i\};n)$ form the full set of genuine
lattice integrals needed to perform one-loop calculations in
lattice perturbation theory.  We are forced to consider $H$ integrals
with $a_i \ne 1$ for two reasons: they appear in the
expansion of loop integrals with external momenta and/or nontrivial
numerators \cite{numer}; in addition, they are needed in the
reduction process of an arbitrary $H$-integral to the master
integrals. 

We now use integration-by-parts identities to fully exploit the
algebraic relations between the  various integrals $H(\{a_i\},n)$. 
These relations are derived using the fact that the integral of a total 
derivative vanishes in analytic regularization:
\begin{equation}
0=  \int \limits_{-\infty}^{\infty} d^{d}\eta
\frac{\partial}{\partial \eta_\mu} \Big \{ \eta_\mu
\prod_{i=1}^d \frac{1}{(1+\eta_i^2)^{a_i}}
\left [ D_B(\eta) \right ]^{-n-\delta}  \Big \} \, ,
\label{recrel}
\end{equation}
for each value of $\mu=1,2,\dots d$.  Computing the derivative in
Eq.(\ref{recrel}) and rewriting the resulting expression in terms of
the functions $H(\{a_i\},n)$, we obtain an algebraic relation between
integrals with different values of $\{a_i\}$ and $n$. Another equation
can be obtained by partial fractioning, i.e.~by using the linear
dependence of the five ``propagators'' in the function $H(\{a_i\},n)$.
The complete set of algebraic relations is therefore
\begin{eqnarray}\label{eq:scalarRel}
0 &=& \bigg\{{\bm n}^- + \sum_{i=1}^{d} ({\bm a}_i^+-1)\bigg\}
H\, , \\
0 &=& \bigg \{ 1 + 2\,a_i\, ({\bm a}_i^+ - 1)   
+ 2\,(n+\delta)\,{\bm n}^+\,{\bm
a}_i^{+}({\bm a}_i^{+}-1)\bigg \} H \, .
\nonumber 
\end{eqnarray}
The conventions are such that the operator ${\bm a}_i^{\pm}$
increases (decreases) the index $a_i$ by one.

Similar integration-by-parts relations for lattice integrals were
first studied in \cite{Caracciolo:1991cp,Burgio:1996ji}, where it was
shown that the entire class of integrals $H(\{a_k\};n)$ can be reduced
to $d$ master integrals in $d$ dimensions. Here, we neither attempt to
solve these equations explicitly nor to rewrite them in such a form
that the reduction of a given index is manifest. Instead, we adopt a
brute force strategy and use computer algebra to explicitly solve the
equations for a given range of indices.  An efficient algorithm for
solving such recurrence relations has been described in
\cite{Laporta:2001dd}. First, a criterion which selects a simpler
integral out of any two integrals is chosen. Typically, integrals with
lower values of the indices are considered to be simpler. The above
equations are then solved for a very limited range of indices, using
Gauss's elimination method. The calculation is repeated after
supplementing the chosen set of equations with a few relations
involving higher index values. By iterating this procedure, the
equations (\ref{eq:scalarRel}) can be solved for the entire index
range needed in a given calculation.  The advantage of this brute
force method is that it immediately generalizes to integrals involving
more complicated propagators (e.g.~those of Wilson fermions) or to
higher loops. In the continuum, this strategy has lead to the solution
of problems that would have been very difficult to tackle by
manipulating recurrence relations by hand, for example the four point
functions of QCD at two loops \cite{Anastasiou:2002zn}.

Having established that the bosonic case requires four master
integrals, we can choose those in any way we like. It is convenient to
choose master integrals that are convergent in the limit $\delta \to 0$, since their expansion in $\delta$ can be obtained by
expanding the integrands in Eq.(\ref{basishard}) in a power series in
$\delta$.  A possible choice that satisfies this criterion is
$H_0=H(\{1,1,1,1\};1)$, $H_1=H(\{1,1,1,2\};1)$,
$H_2=H(\{1,1,2,2\};1)$, $H_3=H(\{1,2,2,2\};1)$.

It is a fairly easy task to numerically evaluate the finite bosonic
integrals to relatively high precision. In the case of $H$
integrals, this task can be further simplified by using the Schwinger
representation for the propagators, which permits a reduction to
one-dimensional integrals. Writing
\begin{equation}\label{eq:Schwinger}
\frac{1}{({\widehat k}^{2})^{\alpha}} =\frac{1}{\Gamma(\alpha)}\,
\int \limits_0^\infty t^{\alpha-1}\,e^{-t\,{\widehat k}^2}\, ,
\end{equation}
the integration over the momentum $k$ factorizes and can be done
analytically, leaving one dimensional integrals over products of
modified Bessel functions $I_n(t/2)$ for numerical evaluation.
In this way, we obtain the following one-dimensional integral
representations for  the master integrals $H_{0-3}$ defined above:
\begin{multline}\label{eq:bosonicbasis}
H_i=\frac{\pi^4}{2^i}\int_0^\infty dt\,e^{-2\,t}\,
{I_0(t/2)}^{4-i}\,\left\{I_0(t/2)+I_1(t/2)\right\}^i\,\\
\times\left[1+(\gamma_E+\ln t)\,\delta\right]+O(\delta^2)\, .
\end{multline}
The precise numerical evaluation of these one dimensional integrals is
straightforward. Our results for the master integrals are
\begin{align}
H_0&= 60.3676829 - 16.7691976\,  \delta  +{\cal O}(\delta^2) \, ,
\nonumber \\
H_1&= 36.0154101 - 1.83118998\, \delta +{\cal O}(\delta^2)\, ,
\\
H_2&= 22.1620171 + 4.75278418\,  \delta +{\cal O}(\delta^2)\, ,
\nonumber\\
H_3&= 14.1541117 + 7.32383339\,  \delta + {\cal O}(\delta^2) \, .\nonumber
\end{align}

Since it is rather simple to obtain the above results, the integrals
$H_{0-3}$ form a convenient basis.  However, in order to facilitate
the comparison of our results with the literature, we point out that
other choices of the master integrals are possible.  One possibility
is 
\begin{align}
\frac{1}{\pi^2}\,H(\{1,1,1,1\};0)&=\pi^2, \nonumber\\
\frac{1}{\pi^2}\,H(\{1,1,1,1\};1)&=4\,\pi^2\,\p{1},\\
\frac{1}{\pi^2}\,H(\{1,1,1,1\};2)&=-\frac{1}{\delta}-2\,\ln 2+1+
16\,\pi^2\,\p{2},\nonumber\\
\frac{1}{\pi^2}\,H(\{1,1,1,1\};3)&=-\frac{1}{2\delta}-\ln 2+
\frac{3}{4}+64\,\pi^2\,\p{3}.\nonumber
\end{align}
The three constants  $\p{1,2,3}$ are closer to traditional choices
of the basis numbers. For instance,  Ref.~\cite{Burgio:1996ji} uses
the notation $Z_0=\p{1}$, $F_0=16\pi^2 \p{2}+\gamma_E$ and
$12\,\pi^2\,Z_1 = 26 + 3\,\pi^2 - 384\,\pi^2\,\p{2} + 3072\,\pi^2\,\p{3}$.
The  two  constants $\p{1}$ and $\p{2}$ are equal to $P_1$ and $P_2$
of \cite{Luscher:1995zz}; however, $\p{3}\neq P_3$.  We can use
the  relations between the two sets of basis integrals to obtain the constants
$\p{1-3}$. For example, the integral $H({1,1,1,1},2)$ is given by
\begin{eqnarray}
 H({1,1,1,1},2) &&=  \frac{(3\delta+5)}{12 (1+\delta)} H_0
+\frac{(3+ 2\delta)}{6(1+\delta)} H_1
\\
&&+\frac{(2\delta+\delta^2-6)}{2(1+\delta)\delta} H_2
+\frac{(2-\delta)^2}{(1+\delta)\delta} H_3. \nonumber
\end{eqnarray}
Using the numerical results for the integrals $H_{0-3}$ presented
above, we then obtain the value of $\p{2}$. Performing similar
calculations for the other integrals, we arrive at
\begin{eqnarray}
\p{1} 
&=&0.15493339,\,\nonumber \\
\p{2} &=&0.02401318,\, \\ 
\p{3} &=&0.00158857\, . \nonumber
\end{eqnarray}

\begin{figure}[htb]
\begin{center}
\includegraphics[width=0.45\textwidth]{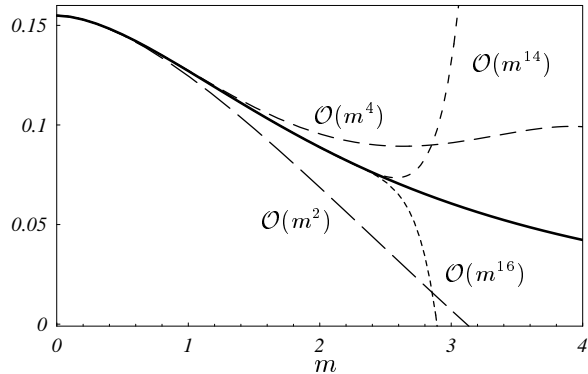}\phantom{abc}
\end{center}
\vspace*{-0.5cm}
\caption{Comparison of the numerical evaluation of the propagator at
the origin $G(m)$ with the expansion around its continuum limit. The thick
line corresponds to the numerical result while the four dashed lines
show the expansions in $m=m_{\rm phys}\,a$ to the indicated
orders. Recall that $m=2\,{\tilde{m}}$.}
\label{fig:numeric}
\end{figure}

Finally, we present the result for the massive lattice 
tadpole-integral by combining the soft and the hard contributions discussed
above.  In Fig.\ref{fig:numeric} we compare the numerical evaluation
of $G(m)$ with the expansion around its continuum limit, including
terms up to $O(m^{16})$. We note that the expansion converges up to
relatively large values of the mass, $m\sim 2.5$, where the Compton
wavelength of the particle is less than half the lattice spacing.

\section{Other propagators}

The scalar lattice propagator in Eq.(\ref{prboson}) arises in the
simplest of infinitely many possible discretizations of the continuum
action. Unfortunately, this simplest choice leads to large
discretization errors.  These can be reduced by introducing improved
actions which include higher dimensional operators chosen to cancel
lattice artifacts to a given order in the expansion around the
continuum limit \cite{Symanzik:1983dc}. In doing so, the free
propagator usually gets modified. A famous example is provided by
fermions on the lattice, where the simplest discretization leads to
the well known doubling problem. Among the many solutions to this
problem, the most common is to follow Wilson's suggestion and add the
dimension-five term $\bar\psi\, \Box\, \psi$ to the fermion action
\cite{wilson}.

Despite the fact that the inclusion of higher dimensional operators in
more refined versions of the lattice action changes the propagator,
our strategy remains the same. The structure of the soft part of the
loop integrals remains unchanged, since it encodes the behavior at
small momenta; the hard part, however, becomes a function of the
couplings of the higher dimensional operators. This typically leads to
more complicated recurrence relations for the hard integrals, but the
approach described in the previous section is robust enough 
to solve the resulting recurrence relations. We now
illustrate this for both static and Wilson fermions.

\subsection{Static Fermions}

In a bound state of a heavy ($m_q\gg \Lambda_{QCD}$) and a light
quark, the heavy quark remains almost static, and the probability of
finding virtual heavy quark-antiquark pairs is small. In this
situation the heavy quark can be treated as an approximately static
color source. The systematic expansion around the static limit is
obtained using the Heavy Quark Effective Theory
(HQET)\cite{Eichten:1989zv}.  This effective theory provides a
method of putting heavy quarks on the lattice that does not require
excessively small lattice spacings.

The discretized heavy quark propagator is \cite{Eichten:1989kb}
\begin{equation}
G_{\rm static}(k_4)=\frac{1}{i(1-e^{i\,k_4})+i\epsilon}\, .
\end{equation}
In the static limit the pole in the propagator occurs at $k_4=0$,
rendering an $i\epsilon$ prescription necessary even when
working in Euclidean space. Using the variables introduced in Section
\ref{sec:notation}, the static propagator can be written as
\begin{equation}
G_{\rm static}(\eta_4)=\frac{1-i\,\eta_4}{2\,\eta_4 +i\,\epsilon}\, .
\end{equation}
The soft integrals are obviously the ordinary continuum HQET integrals.
After regulating the bosonic propagator, the hard integrals become
\begin{eqnarray}
H(\{a_i\};n,m)=\int \limits_{-\infty}^{\infty}
\prod_{i=1}^d \frac{d\eta_i}{(1+\eta_i^2)^{a_i}}
\frac{[ D_B(\eta)]^{-n-\delta}}{(\eta_4+i\,\epsilon)^m}\, .
\label{basishardStatic}
\end{eqnarray}

Again, we wish to use integration-by-parts relations to reduce the
number of independent hard integrals to an absolute minimum. Both the relations obtained by differentiation with respect to
$\eta_{1-3}$ and the partial fractioning identity for the bosonic
propagator coincide with the relations given in the previous section.
The integration-by-parts identity obtained by differentiating with
respect to $\eta_4$ becomes
\begin{eqnarray}
&& \left \{ 2\left(n+\delta\right)\,\left({\bm a}_1^+-{\bm m}^{++}\right)
\, {\bm a}_1^+\, {\bm n}^+ 
\right. \nonumber \\
&& \left. - \left [ \left(3+m\right)\,{\bm m}^{++}
+ 2\,a_1\, {\bm a}_1^+\right ] \,{\bm m}^{++}\right \} H=0 \, .
\end{eqnarray}
The partial fractioning
identity for the static propagator is
\begin{equation}
\left\{\left(1-{\bm a}_1^+\right)\,{\bm m}^{++} - {\bm a}_1^+ \right\} 
H=0\,.
\end{equation}
An additional algebraic identity is obtained by replacing
\begin{equation}
\frac{1}{\eta_4+i\,\epsilon} = {\cal P}\,\frac{1}{\eta_4} 
- i\,\pi\,\delta(\eta_4) \,.
\end{equation}
The principal value contribution to $H(\{a_i\};n,1)$ vanishes 
because of the symmetry $\eta_4 \to -\eta_4$,  and the
contribution from the $\delta$-function is independent of $a_4$:
\begin{equation}
H(\{a_1,a_2,a_3,a_4\};n,1)=H(\{a_1,a_2,a_3,0\};n,1)\, .
\end{equation}
Performing the loop integration over $\eta_4$, it is easy to see that
the integrals $H(\{a_1,a_2,a_3,0\};n,1)$ reduce to the bosonic hard
integrals in three dimensions.  The lattice HQET master integrals are
therefore given by the four bosonic master integrals discussed in the
previous section, and the three bosonic master integrals in $d=3$
\cite{d3}. For the three-dimensional bosonic integrals we choose the
basis
\begin{align}
\frac{1}{\pi^3}\,H(\{1,1,1\},0)&=1+{\cal O}(\delta) \,, \nonumber \\
\frac{1}{4\,\pi^3}\,H(\{1,1,1\},1)&=\q{1}+{\cal O}(\delta)\approx
0.252731010 \,, \\ \frac{1}{16\,\pi^3}\,H(\{1,1,1\},2)&=\q{2}+{\cal
O}(\delta)\approx 0.012164159 \,.  \nonumber
\end{align}

\subsection{Wilson fermions}

Wilson fermions \cite{wilson} are often used to introduce fermions on
the lattice, and we therefore consider this case in detail. The
propagator of a Wilson fermion is
\begin{equation}
G(k)=\frac{-i\,\sum_\mu
\gamma_\mu\,\sin(k_\mu)+m+\frac{r}{2}\,{\widehat{k}}^2 }{4\,D_F(k,m)}\, .
\end{equation}
For our discussion only the form of its denominator,
\begin{equation}
4\,D_F(k,m)=\sum_\mu \sin^2(k_\mu)+(m+\frac{r}{2}\,{\widehat{k}}^2)^2\, ,
\end{equation}
is relevant. In the continuum limit, external momenta and masses tend
to zero while the Wilson parameter $r\neq 0$ stays fixed. In the
notation of Section \ref{sec:notation} the denominator is
\begin{eqnarray}
D_F(\eta,\tilde m) =\sum_\mu \frac{\eta_\mu^2}{(1+\eta_\mu^2)^2}+(\tilde m
+r\, D_B(\eta))^2 \,.
\end{eqnarray}
We consider a general one loop integral with both bosonic and
fermionic propagators and regulate the fermionic propagator. The
integrals that appear in the hard part of the expansion are
\begin{eqnarray}
&& H\equiv H(\{a_i\},n,m) =
\\
&& \int {\rm d}^d \eta \prod_{i=1}^d
\frac{1}{(1+\eta_i^2)^{a_i}}\,
D_B(\eta)^{-n}~D_F(\eta,0)^{-m-\delta}.
\nonumber 
\end{eqnarray}

We again begin by writing down the algebraic and integration-by-parts 
identities for the hard integrals. We find three identities
by partial fractioning:
\begin{eqnarray}
&&\bigg\{1+\sum_i ({\bm a}_i^+ - 1) {\bm n}^+ \bigg\}\,H =0, \\
&&\bigg\{1+\sum_i 
({\bm a}_i^+ - 1)\,{\bm m}^+
\nonumber \\
&&
~~~~~~~~~~~-(r^2-1) 
\Big [ \sum_i({\bm a}_i^+ - 1)\, \Big ]^2 {\bm m}^+\bigg\}\,H
=0,\nonumber \\
&&\bigg\{\Big[{\bm m}^- + \sum_i {\bm a}_i^+\,({\bm a}_i^+ -
1)\Big]\,{\bm n}^{++} - r^2 \bigg\}\,H =0; \nonumber
\end{eqnarray}
$d$ additional ones arise from integration-by-parts:
\begin{multline}
\bigg\{
1 + 2\,(m+\delta)\, {\bm a}_i^+ ({\bm a}_i^+-1) {\bm m}^+ \Big[ 2\,{\bm
a}_i^+ -1 - 2\, r^2 \sum_j ({\bm a}^+_j-1)\Big] \\
 +2\,a_i\,({\bm a}_i^+ - 1)+2\,n\, {\bm a}_i^+\, ({\bm
a}_i^+-1)\, {\bm n}^+ 
  \bigg\} H=0 \, .
\end{multline}
These equations are considerably more complicated than the ones
encountered in the scalar case. Not only does the index space become
six-dimensional, but the recurrence relations also shift indices by up
to three units.  While the solution of the bosonic recursion relations
in the index range needed for the applications of Section
\ref{sec:applications} was accomplished within hours, about a day of
computer time was required to perform the reduction in the fermion case
for $r=1$. We are left with sixteen master integrals, which must
be computed numerically. For purely fermionic integrals this number
reduces to ten.

Because of the complicated structure of the fermion denominator, we
cannot use the Schwinger representation (\ref{eq:Schwinger}) to arrive
at one-dimensional representations for the master integrals. A general
strategy to obtain accurate results for more complicated cases is to
choose a basis of integrals that are very convergent in the
infrared, such as those with negative powers of
propagators. Because of the simple behavior of the integrand at small
values of $k$, these integrals can be obtained by replacing
the integration with a sum over a small lattice. We illustrate
this point for the purely bosonic integrals. Instead of the four integrals
$H_i$ defined in Eq.~(\ref{eq:bosonicbasis}), we could have chosen
to evaluate four integrals from the set
\begin{equation}
H(\{1,1,1,1\},-n)=
\int_{-\pi}^{\pi} d^4 k\, \left(\frac{{\widehat{k}}^2}{4}\right)^{n+\delta}
\end{equation}
to second order in $\delta$. Replacing $k_i\rightarrow
\frac{2\,\pi}{L}(l_i-\frac{1}{2})$ and trading the integral over
the Brillouin zone for a sum over $l_i=1,\dots, L$, we obtain the
value of the above integral up to finite size effects of order
$1/L^{4+2n}$. Even the choice $n=0,1,2,3$ gives rather accurate values
for lattice sizes $L>50$.  In reference \cite{Burgio:1996ji}, this
strategy was applied to fermionic tadpole-integrals and precise values for
the corresponding master integrals were derived. We have verified the
first few digits of their numerical results.

\section{Applications\label{sec:applications}}

\begin{figure}[htb]
\begin{tabular}{cc}
\includegraphics[width=30mm]{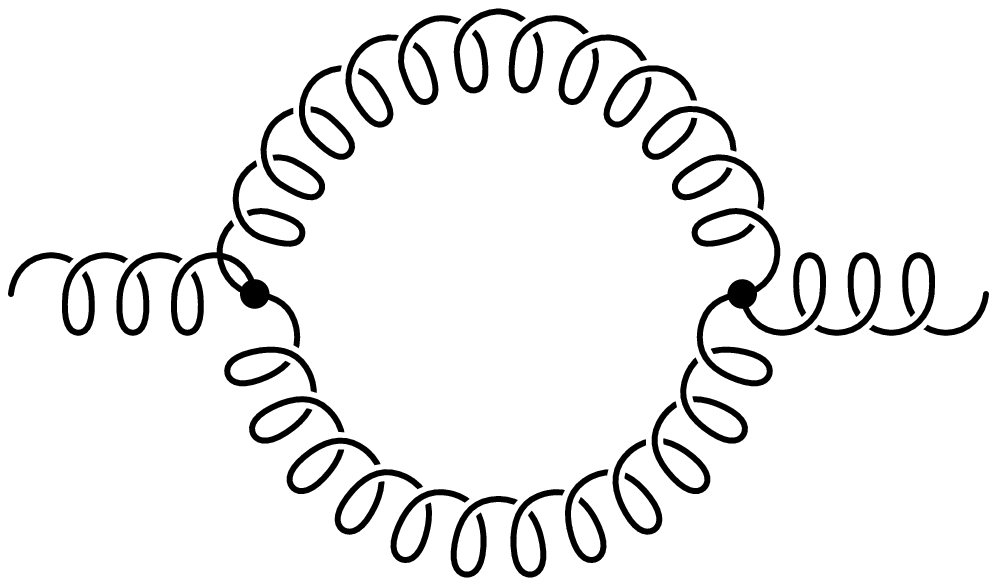}
&\hspace*{0mm}
\includegraphics[width=30mm]{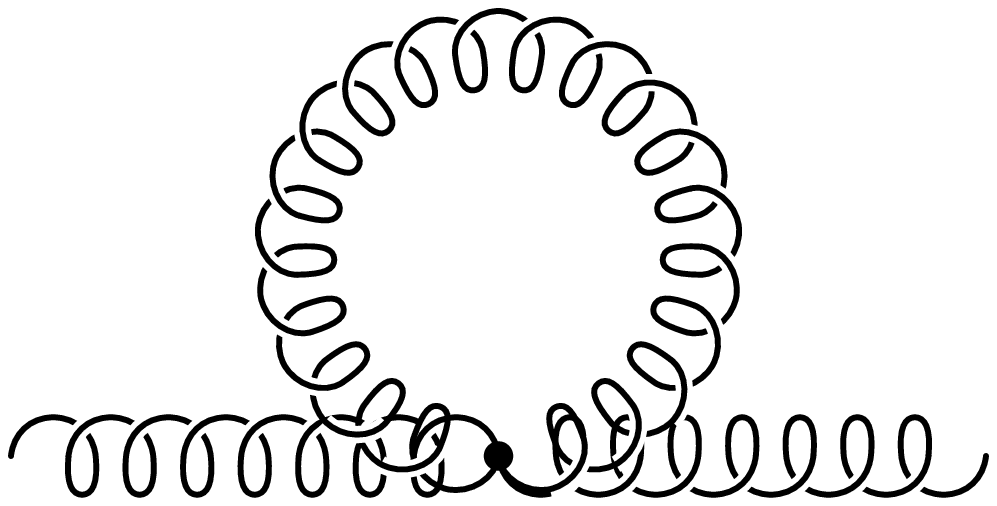}
\\[1mm]
(a) & (b)\\
\\[1mm]
\includegraphics[width=30mm]{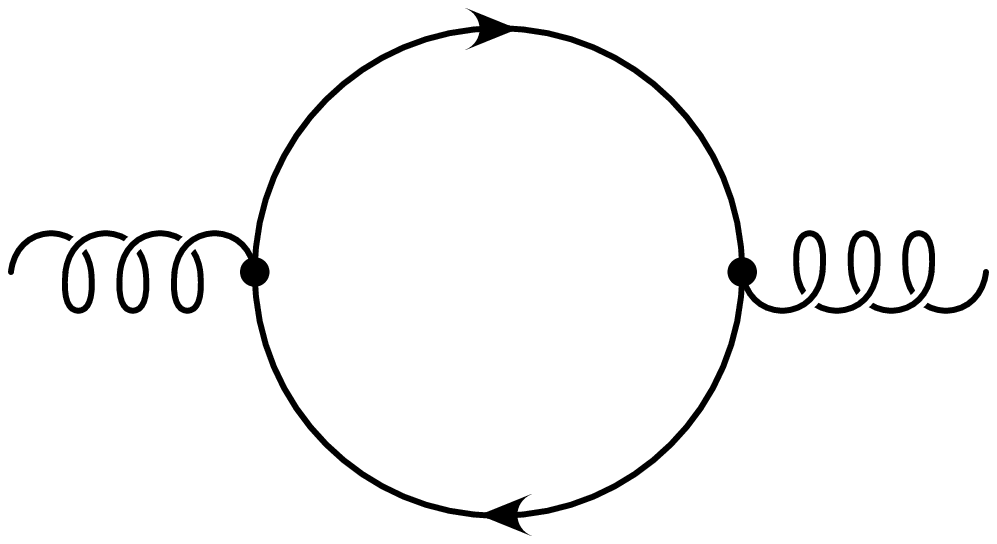}
&\hspace*{0mm}
\includegraphics[width=30mm]{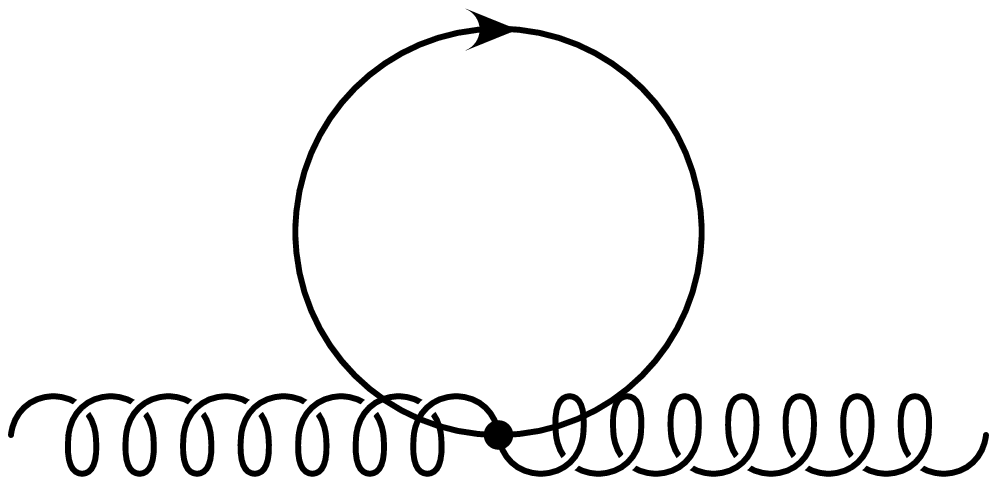}
\\[1mm]
(c) & (d)\\
\\[1mm]
\includegraphics[width=30mm]{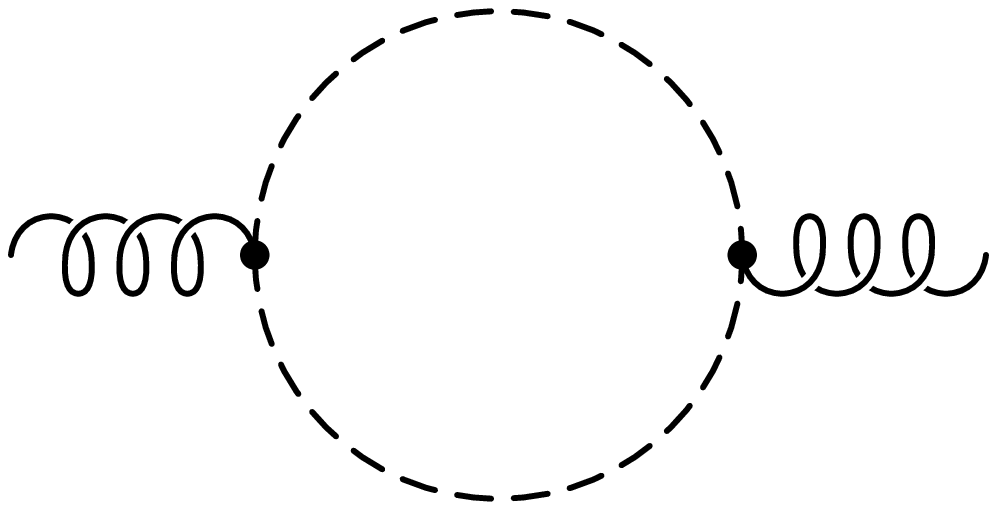}
&\hspace*{0mm}
\includegraphics[width=30mm]{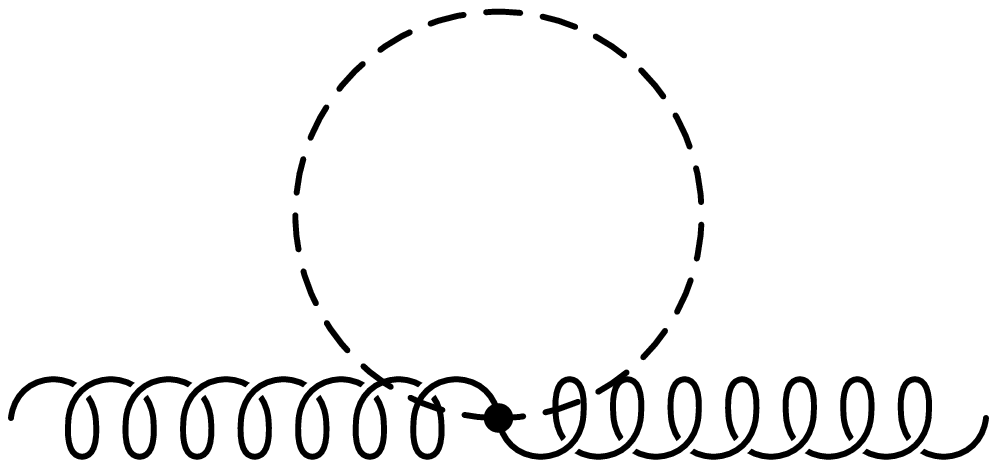}
\\[1mm]
(e) & (f)\\
\\[1mm]
\multicolumn{2}{c}{\includegraphics[width=30mm]{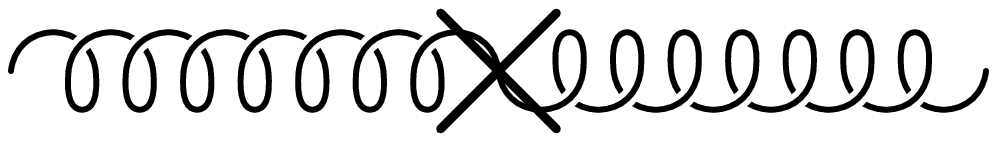}}
\\
\multicolumn{2}{c}{(g)}
\end{tabular}
\caption{Feynman diagrams contributing to the gluon self energy at
$O(g^2)$.  The graphs {\it d}, {\it f} and {\it g} have no continuum
analog. The graph {\it g} arises from the expansion of the Haar
measure.}
\label{fig:graphs}
\end{figure}

To test this method in a practical application, we use it to
rederive the known results for the gluon and fermion self-energies in
QCD.  The calculation of these quantities is instructive since it
exhibits all the complications that arise in lattice perturbation
theory.  For example, the expansion of the gauge action around the
continuum limit leads to infinitely many vertices. When they multiply
divergent integrals, sub-leading vertices contribute in the continuum
limit and lattice calculations consequently involve a larger number of
graphs, than the corresponding calculations in the continuum
theory. Furthermore, the vertices in lattice gauge theory have a
complicated form. It takes roughly one page to write down the Feynman
rule for the four gluon vertex \cite{Rothe:kp}. The tensor structure
of the graphs can therefore be quite lengthy. Summing over the
internal indices in the first graph in Fig.~\ref{fig:graphs} one ends
up with several hundred terms. However, the large number of terms is
not a significant obstacle since the technique described here can be easily automated.

\subsection{Gluon self-energy}
The individual gluon self-energy graphs of Fig.~\ref{fig:graphs}
diverge quadratically with the lattice spacing, and we have
therefore expanded them to sub-leading order.  They contain terms
proportional to the the lattice invariants $g_{\mu \nu}$,
$p_\mu\,p_\nu$ and $p_\mu^2\,g_{\mu\nu}$. In the sum the quadratic
divergences and the non-covariant pieces cancel and the result for
the inverse gluon propagator becomes
\begin{multline}
\Gamma^G_{\,\mu\nu,a b}(p)=
\left(g_{\mu,\nu}-p_{\mu}\,p_{\nu}\right)\,\delta_{a b}\, 
\bigg[
1+ \frac{g^2}{(4\pi)^2} 
\big(
\Gamma_{\rm hard}
 \\ 
+\Gamma_{\rm soft}\big) \bigg ]
+(\xi+1)\,\delta_{a b}\,p_\mu\,p_\nu.
\end{multline}
The gluonic contribution to the self-energy is
\begin{multline}
\Gamma^{(G)}_{\rm soft} =N\bigg\{-
\frac{5}{3\,\delta } -\frac{14}{9} - \frac{{\xi }^2}{4} +
\frac{5}{3}\,\ln\frac{p^2}{4} \\
+ \xi \,\left(\frac{1}{2\,\delta } -
\frac{5}{6}  - \frac{1}{2}\ln\frac{p^2}{4} \right)\bigg\} \, , \\
\Gamma^{(G)}_{\rm hard}=N\bigg\{ \frac{5}{3\,\delta }  -
  \frac{14}{9} - {\pi }^2 + \frac{2\,{\pi }^2}{N^2} -
  \frac{14\,{\pi }^2}{9}\,\p{1} -
  \frac{80\,{\pi }^2}{3}\,\p{2} \\ + \frac{5}{3}\,\ln 4   +
  \xi \,\left(- \frac{1}{2\,\delta } - \frac{1}{6} -
  \frac{{\pi }^2}{3}\,\p{1} + 8\,{\pi }^2\,\p{2}  - \ln 2
\right)\bigg\}\nonumber \, ,
\end{multline}
where $N=3$ is the number of colors. The infrared divergence
at $\delta=0$ in the hard part cancels the ultraviolet
divergence of the soft part after the two terms are added.  For $r=1$,
the Wilson fermion contribution to the gluon vacuum polarization is
\begin{eqnarray}
&& \Gamma^{(f)}_{\rm hard} =
\frac{1}{2}\Big\{\frac{4}{3\,\delta} -1.376\Big\} \, ,
\nonumber\\
&&\Gamma^{(f)}_{\rm soft}  =
\frac{1}{2}\Big\{-\frac{4}{3}\,L_\delta-\frac{10}{9} + \frac{16\, x}{3} 
+ \frac{4}{3}\,\left( 1-2\,x \right)
\,f(x)
\Big\}, \nonumber 
\end{eqnarray}
where $x=m^2/p^2$, $L_\delta=1/\delta - \ln(m^2/4)$, and
$$
f(x)=\sqrt{1+4\,x} \ln\frac{\sqrt{1+4\,x}+1}{\sqrt{1+4\,x}-1}.
$$
After combining the hard and soft parts, we reproduce the result of
Refs.~\cite{Hasenfratz:1980kn,Kawai:1980ja}.

\subsection{Wilson fermion self-energy}
For $r=1$, the self-energy of a Wilson fermion with mass $m$ is
\begin{align}
\Sigma(p)&=\frac{C_F\,g^2}{( 4\,\pi)^2}\,\left(\Sigma^{(0)}+i\,
p\!\!\!/\,\Sigma^{(1)}+ m\,\Sigma^{(2)}\right) \, .
\end{align}
The hard part is given by
\begin{align}
\Sigma^{(0)}_{\rm hard}&=-51.435\, , \nonumber \\
\Sigma^{(1)}_{\rm hard}&=\frac{1+\xi}{\delta}+13.74-2.91\,\xi  \,,
\nonumber\\
\Sigma^{(2)}_{\rm hard}&=\frac{4+\xi}{\delta}+3.45-3.41\,\xi \,;\nonumber
\end{align}
the soft part is
\begin{eqnarray}
&& \Sigma^{(0)}_{\rm soft}=0 \nonumber \, ,\\
&& \Sigma^{(1)}_{\rm soft}=(1 + \xi)\, 
\Big\{-L_\delta + x - \frac{1}{2} 
+ \left(1 -  x^2 \right) \,\ln\frac{1+x}{x} \Big \} \, , \nonumber\\
&& \Sigma^{(2)}_{\rm soft} =\left ( 4 +
\xi \right) \,\Big\{-L_\delta  
-1 + ( 1+ x)\,\ln\frac{1+x}{x}
 \Big \} \, .\nonumber 
\end{eqnarray}
with $x=m^2/p^2$, $L_\delta=1/\delta - \ln(m^2/4)$.

\subsection{Static quark self-energy}

The wave function renormalization of the heavy quark is infrared
divergent. Regulating the infrared divergence with a gluon mass
$\lambda$, the self-energy takes the form
\begin{equation}
\Sigma_{\rm HQET}(\omega) = \frac{C_F\,g^2}{(
4\,\pi)^2}\,\left(\Sigma_{\rm hard}
+\Sigma_{\rm soft}\right),
\end{equation}
where ($x = \lambda^2/\omega^2,~L_\delta=1/\delta -\ln(\lambda^2/4)$) and
\begin{eqnarray}
 \Sigma_{\rm hard} &=&-8\,i\,\pi^2\,\q{1}+\omega\,\bigg\{- \frac{\left( 2 - \xi
   \right) }{\delta } + 8\,\pi^2\,\q{1} -2\,{\p{1}}\,\pi^2 
\nonumber \\
&& 
+ 16\,{\p{2}}\,\pi^2\,\left(2 - \xi \right) 
-  \xi - 2\,\left( 2 - \xi \right)
   \,\ln 2 \bigg\},
\nonumber\\
 \Sigma_{\rm soft} &=& \omega\, \bigg \{
  \left( 2 - \xi \right)L_\delta
   + 4+\xi
 \nonumber\\
&&+ \frac{\left( 2\, x + \left( 2 - \xi
   \right) \right) }{{\sqrt{1 + x }}}\, 
   \left( i \,\pi + \ln \frac{\sqrt{1+x} - 1
   }{\sqrt{1 + x} + 1}  \right)\bigg \}.\nonumber
\end{eqnarray}

Upon extracting the mass and wave-function renormalization from the above
expression, we recover the result of Ref.~\cite{Eichten:1989kb}.

\section{Conclusions}

We have presented a simple technique for calculating the expansion of
lattice integrals around their continuum limit.  The method is based
on two observations. First, using techniques developed for continuum
integrals, it is possible to systematically expand the lattice
integrals in a series in the lattice spacing. The expansion is
non-analytic and requires introducing an intermediate regularization
to make the expansion of the integrands possible.  After the
expansion, the original integral is a sum of two distinct
contributions.  The first one is a set of analytically regularized
continuum integrals which encode the small momentum behavior of
the original lattice integral and depend in a non-trivial way on the
internal momenta and masses.  The remaining part is a polynomial in
the momenta and masses with massless lattice tadpole-integrals as
coefficients.  Within a given theory, these tadpole-integrals are
process-independent and need to be calculated only once.

We have shown how to efficiently calculate massless lattice tadpoles.
Integration-by-parts identities relate different lattice 
tadpole-integrals.  These relations are rather complicated, and their analytic
solution is difficult.  However, it is not necessary. A robust
approach is to use the algorithm suggested in \cite{Laporta:2001dd}
which allows the solution of the recurrence-relations and the
reduction of all the lattice tadpole-integrals to a few master
integrals to be performed in an automated fashion.  We have applied
this algorithm to tadpole-integrals in gluodynamics, HQET on the
lattice and QCD with Wilson fermions. We have illustrated the
flexibility of the method by computing a number of two-point functions
in lattice perturbation theory such as the gluon and fermion
self-energies and the static fermion self-energy in an arbitrary
gauge.

In the future, it will be interesting to see if this method can be
generalized to higher orders in perturbation theory.  It is clear that
a similar separation of the integrals into process-dependent soft and
universal hard parts is possible;  it should also be possible
to derive integration-by-parts relations for the hard parts of the lattice
integrals.  A potential problem could arise from the size and the
complexity of the system of equations governing the reduction of the
massless tadpole-integrals at the two-loop level. In principle, the
size of the system is not very relevant, but for practical reasons,
such as available CPU time and memory requirements, it might impose
 severe constraints. 

Although it may sound paradoxical, more precise data from lattice
simulations will not eliminate the need for perturbation theory. In
many cases the nonperturbative results from simulations need to be
matched by equally precise perturbative calculations in lattice
regularization in order to become phenomenologically relevant. Using
the method discussed in this paper, such calculations are not much more
difficult than perturbative calculations in continuum field
theory. Hopefully, this simplification will persist at higher orders.

\noindent{\bf Acknowledgments:} We are grateful to Frank Petriello for useful
suggestions and to Charalampos Anastasiou for sharing some of his
insight in solving recurrence relations by computer algebra. This
research was supported by the DOE under grant number
DE-AC03-76SF00515.


\begin{thebibliography}{99}
\bibitem{d3} We note that the hard master integrals in two dimensions
can be evaluated analytically: $H(\{1,1\};1)=-\pi/\delta+3\,\pi\,\ln
2+O(\delta)$ and $2\,H(\{1,1\};2)=-\pi/\delta-\pi+3\,\pi\, \ln
2+O(\delta)$.
\end{thebibliography}
\end{document}